\documentclass[11pt, letterpaper]{article}
\usepackage[left=1in, right=1in, top=1in, bottom=1in]{geometry}
\usepackage[utf8]{inputenc}
\usepackage[T1]{fontenc}
\usepackage{bm}
\usepackage{type1cm}
\usepackage{lettrine}
\usepackage{amsmath,amssymb,amsthm}
\usepackage{moreverb}
\usepackage{mathtools}
\usepackage{amsmath}
\usepackage{amssymb}
\usepackage{algorithmic}
\usepackage{graphics}
\usepackage{graphicx}
\usepackage{caption}
\usepackage{extarrows}
\usepackage{color}
\usepackage{framed}
\usepackage{wrapfig}
\usepackage{bm}
\usepackage{mathrsfs}
\usepackage{mathabx}
\usepackage{multirow}
\usepackage{longtable}
\usepackage{hyperref}
\usepackage{paralist}
\usepackage{indentfirst}
\usepackage{relsize}
\usepackage{extarrows}
\usepackage{lineno}
\usepackage{upgreek}
\usepackage{bm}
\usepackage{mwe}
\usepackage[dvipsnames,table]{xcolor}
\usepackage{booktabs}
\usepackage{authblk}
\usepackage{subcaption}
\usepackage{lettrine}
\usepackage{type1cm}
\usepackage{threeparttable}
\usepackage[sort&compress,numbers]{natbib}
\usepackage[figurename=Figure]{caption}
\usepackage[normalem]{ulem}
\usepackage{tcolorbox}
\usepackage{pgfplots}
\usepackage{arydshln} 
\usepackage{enumitem}
\usepackage{pifont}
\definecolor{yesgreen}{RGB}{0,140,0}
\definecolor{nored}{RGB}{200,0,0}
\usepackage{siunitx}
\usepackage{makecell}
\usepackage{placeins}    
\usepackage{float}       

\graphicspath{ {./Figure/} }
\usepackage[font=footnotesize,labelfont=bf]{caption}

\providecommand{\keywords}[1]{\textbf{\textit{Keywords: }} #1}

\hypersetup{
bookmarks=true,
bookmarksopen=true,
bookmarksnumbered=true,
unicode=false,
pdftoolbar=true,
pdfmenubar=true,
pdffitwindow=false,
pdfstartview={FitH},
pdftitle={My title},
pdfauthor={Author},
pdfsubject={Subject},
pdfcreator={Creator},
pdfproducer={Producer},
pdfkeywords={keywords},
pdfnewwindow=true,
colorlinks=true,
linkcolor=blue,
citecolor=blue,
filecolor=blue,
urlcolor=blue
}

\definecolor{natureDarkBlue}{RGB}{34, 84, 126}
\definecolor{natureMidBlue}{RGB}{68, 149, 212}
\definecolor{natureLightBlue}{RGB}{189, 215, 238}
\definecolor{natureGray}{RGB}{235, 235, 235}

\pgfplotsset{
    nature_style/.style={
        font=\sffamily\scriptsize,  
        width=6.0cm, height=5.5cm,  
        axis line style={draw=none}, 
        axis x line*=bottom,
        axis y line*=left,
        axis line style={black!80, line width=0.8pt},
        ymajorgrids=true,
        grid style={natureGray, dashed},
        tick align=outside,
        tick style={black!50},
        legend style={draw=none, fill=none}
    }
}

\definecolor{purple}{HTML}{c994c7}
\definecolor{navyblue}{RGB}{30,130,255}
\definecolor{citecolor}{RGB}{30,130,255}
\definecolor{lightgray}{gray}{0.9}
\definecolor{blanchedalmond}{rgb}{1.0, 0.92, 0.8}
\definecolor{cerise}{rgb}{0.871, 0.192, 0.388}

\definecolor{TaskBG}{HTML}{EFE6FF}        
\definecolor{StateBG}{HTML}{F5F5F7}       
\definecolor{ExpertBG}{HTML}{EAF7EA}      
\definecolor{IWMBG}{HTML}{FDECF3}         
\definecolor{SRBG}{HTML}{E6F2FF}          

\newtcolorbox{trainingexample}[2][]{%
  enhanced, breakable, colframe=black!12, colback=white, boxrule=2.5pt,
  arc=2pt, left=0pt, right=0pt, top=0pt, bottom=0pt,
  title={#2}, fonttitle=\bfseries, coltitle=black, #1}

\newcolumntype{L}[1]{>{\RaggedRight\arraybackslash}p{#1}} 
\newcolumntype{Y}{>{\RaggedRight\arraybackslash}X}        

\AtBeginDocument{%
  }

\definecolor{lemon}{HTML}{FDFFCC}

\definecolor{Gainsboro}{rgb}{0.86, 0.86, 0.86}

\definecolor{Gray}{gray}{0.95}
\definecolor{LightCyan}{rgb}{0.88,1,1}
\definecolor{dm-blue-500}{RGB}{0, 69, 177}
\definecolor{dm-purple-500}{RGB}{105,50,230}
\definecolor{dm-red-500}{RGB}{255,122,122}

\definecolor{backred}{RGB}{255, 190, 190}
\definecolor{backblue}{RGB}{220, 230, 250}

\newtcbox{\hlprimarytab}{on line, rounded corners, box align=base, colback=backblue, colframe=white, size=fbox, arc=3pt, before upper=\strut, top=-2pt, bottom=-4pt, left=-2pt, right=-2pt, boxrule=0pt}
\newtcbox{\hlsecondarytab}{on line, box align=base, colback=backred, colframe=white, size=fbox, arc=3pt, before upper=\strut, top=-2pt, bottom=-4pt, left=-2pt, right=-2pt, boxrule=0pt}

\begin{document}
\title{\textbf{Benchmarking LLMs for Community Governance Simulation with Life-history Narratives}}

\author[1,$\dag$]{Xu Chen}
\author[1,$\dag$]{Yuanzi Li}
\author[1]{Lei Wang}
\author[2]{Nan LU}
\author[2]{Yang Wang}
\author[3]{Anding Wang}
\author[2,$^*$]{Lei Shi}
\author[2,$^*$]{Xiaoxing Fu}
\author[1,$^*$]{Ji-Rong Wen}

\affil[1]{\small Gaoling School of Artificial Intelligence, Renmin University of China, Beijing, China}
\affil[2]{\small School of Social Research, Renmin University of China, Beijing, China}
\affil[3]{\small Big Data and Responsible Artificial Intelligence for National Governance, Renmin University of China, Beijing, China}
\affil[$\dag$]{Co-first authors}
\affil[*]{Corresponding authors}

\date{}

\maketitle

\normalsize

\vspace{-18pt}
\begin{abstract}
Effective community governance hinges on understanding what specific residents think and need. Recent work has used large language models (LLMs) to simulate human respondents, offering a scalable, reproducible way to study human attitudes and behaviors at low cost. However, these studies typically prompt the model with just a few demographic variables (age, gender, income), simulating only general role types. This is insufficient for community governance, where decisions depend on the views of specific residents. We bridge this gap with an integrated research framework covering dataset, benchmark, algorithm, and system. The dataset comprises approximately 1.2 million characters of first-person narrative collected through two-hour semi-structured interviews with each of 92 residents in an urban community, organized around nine community-governance domains. The benchmark probes 18 mainstream LLMs across four prompting strategies and shows that adding rich life-history profiles meaningfully raises fidelity above the no-profile baseline, but this gain comes with more input tokens per call from the longer prompts they require. The algorithm, \textsc{curriculum-LoRA}, is a parameter-efficient personalization framework that, by closing this fidelity--cost gap, matches the strongest baseline's fidelity at roughly $10\times$ lower per-call cost and Pareto-dominates every configuration tested. The system integrates \textsc{curriculum-LoRA} into a closed-loop policy-evaluation pipeline. Together, these results bring individual-level LLM-based resident simulation within reach of resource-constrained local administrations, enabling community-governance decisions to be systematically pre-evaluated {in silico} before real-world deployment.
\end{abstract}


\vspace{12pt}
\section*{Introduction}
Community governance is the foundation of social stability and public trust. As the most local scale of public administration, it is where policies on service provision, resource allocation, and conflict resolution are translated into residents' lived experiences~\cite{ostrom1990governing,putnam2000bowling}. The effectiveness of governance at this level depends critically on understanding what residents actually think, need, and are likely to do in response to specific proposals~\cite{soss2000unwanted,schneider1993social}. Without such understanding, policies risk misalignment with local realities, eroding public trust and undermining social cohesion~\cite{sampson2012great,heberer2009community}.
Acquiring this knowledge, however, remains remarkably difficult. Collecting high-quality responses at scale is slow and expensive: each respondent must be recruited, compensated and engaged for substantial time, which puts routine, broad-coverage data collection beyond the reach of most resource-constrained local governments~\cite{groves2011survey,dillman2014internet}. Perhaps more fundamentally, conventional surveys are inherently one-shot: the same respondents cannot be re-queried under counterfactual policy designs, leaving decision-makers unable to evaluate proposals before real-world deployment~\cite{salganik2019bit,manning2024automated}.

The recent emergence of large language models (LLMs) offers a way to relax both constraints: they are cheap to query, reproducible, and re-runnable. A fast-growing body of work shows that, when prompted with demographic or attitudinal profiles, LLMs can reproduce human behavioural tendencies across domains ranging from political science to urban planning~\cite{argyle2023one,santurkar2023whose,lin2026alignsurvey,aher2023using,horton2023large,park2023generative,wang2025user,zhang2025simulating}. Yet these studies almost uniformly prompt the model with sparse, generic persona descriptions: a handful of demographic variables such as age, gender, income and education, or brief synthetic biographies assembled without direct contact with real individuals~\cite{hwang2023aligning,jiang2023personallm}. Such inputs let an LLM at most play a demographic \emph{role type}, not any \emph{specific} individual. Community governance, however, demands more than this. As the most local level of public administration, community work deals with residents on a person-by-person basis, and what matters is knowing exactly which individuals will react in which ways, not group averages. A single strong objector can turn neighbours against a policy; a single overlooked household can become a lasting source of conflict; a single key supporter can win over a hesitant group. Capturing this heterogeneity requires individual-level simulation, which demands rich per-resident profiles to distinguish different individuals. However, feeding large profiles into commercial LLMs is expensive, which conflicts with the tight budgets of community governance and the large number of simulations it typically requires. Therefore, the central challenge of resident simulation is the trade-off between fidelity and cost.

To the best of our knowledge, no prior work has attempted to drive LLMs with large-scale, fine-grained resident profile data for community governance. To lay the foundation for this direction and address the gaps above, we propose an integrated research framework with four components: dataset, benchmark, algorithm, and system. \emph{First}, the \emph{dataset}: life-course research holds that individual attitudes are shaped progressively by a person's trajectory of life events and how those events are subjectively interpreted~\cite{elder1998life,mayer2009new}. We therefore collected each resident's life-history together with their policy attitudes. We conducted semi-structured interviews~\cite{guest2006many} of approximately two hours each with 92 long-term residents in a representative urban community. The resulting corpus has two complementary components: (i) a life-history narrative per resident, organized into four parts covering basic information and growth; education, work and migration history; marriage, family and care; and community interaction and personal values; and (ii) a structured policy-attitude record from a 50-question instrument covering nine core dimensions of community governance. Across the 92 residents, this yields approximately 1.2 million characters of first-person narrative. Because each resident is now represented by a rich life-history, this corpus enables, for the first time to our knowledge, systematic research on more individualized LLM role-play in the community-governance domain. \emph{Second}, the \emph{benchmark}: we systematically evaluated 18 mainstream LLMs that span open-source families such as Qwen~\cite{bai2023qwen}, GLM~\cite{glm2024chatglm}, and Kimi~\cite{team2025kimi}, and proprietary frontier models from OpenAI~\cite{achiam2023gpt}, Google~\cite{team2023gemini}, and other leading commercial providers. Each was combined with four representative prompting strategies (zero-shot~\cite{brown2020language}, life-history~\cite{deshpande2023toxicity, argyle2023out, morocho2026assessing}, life-history-free few-shot~\cite{brown2020language}, and life-history-augmented few-shot), yielding 72 model--prompt configurations. We find that adding rich life-history profiles to the prompt meaningfully raises individual-level simulation fidelity above the no-profile baseline; however, the best prompting configurations plateau near 50\% accuracy, and they cost roughly an order of magnitude more per call than mid-scale alternatives that come within a few percentage points of them. The accuracy--cost trade-off offered by pure prompting is therefore sharply unfavourable, and no model swap or shot tuning closes it. \emph{Third}, the \emph{algorithm}: to close this gap, we propose \textsc{curriculum-LoRA}, a parameter-efficient personalization framework that augments persona-conditioned supervised fine-tuning~\cite{ouyang2022training,suh2025language,krsteski2025valid} with two key components: online random sampling of \emph{reference questions} that lets the model exploit cross-question consistency in each resident's preferences, and a curriculum~\cite{bengio2009curriculum,wang2021survey,soviany2022curriculum} that gradually increases the number of reference questions during training so that the model first learns the basic persona-to-answer mapping before leveraging richer cross-question context. Built on a 7B-parameter base model, \textsc{curriculum-LoRA} matches the strongest prompting baseline's fidelity (51.6\% vs GLM-5's 49.7\%) while reducing per-call inference cost by an order of magnitude or more (CNY 0.41 for all evaluation calls, against CNY 6.54 for GLM-5, CNY 13.84 for GPT-5 and CNY 38.12 for GPT-4.1 under the same prompting strategy). On the joint accuracy--cost plane it Pareto-dominates~\cite{pareto1919manuale} every other configuration in our benchmark, and it further generalizes to unseen residents and unseen governance domains. \emph{Fourth}, the \emph{system}: we integrate the calibrated model into an end-to-end policy-attitude simulation system that supports a complete governance workflow: policy-probe design, resident-attitude simulation, result analysis, and iterative policy-probe refinement. This closes the loop from policy hypothesis to in-silico evaluation and back.

In summary, this work makes three contributions. \emph{First}, we construct the largest individual-level resident-simulation benchmark to date, comprising approximately 1.2 million characters of life-history narrative collected through roughly two-hour semi-structured interviews with each of 92 residents, paired with their structured governance responses across nine dimensions. \emph{Second}, we systematically map the accuracy--cost frontier of mainstream LLMs on this benchmark, showing that pure prompting offers a sharply unfavourable accuracy--cost trade-off that precludes practical deployment. \emph{Third}, we develop \textsc{curriculum-LoRA}, a parameter-efficient personalization framework that Pareto-dominates every prompting configuration tested by matching their fidelity at an order of magnitude lower per-call cost, and we integrate it into a closed-loop policy-evaluation system that practitioners can actually use.

\begin{figure*}[!tb]
  \centering
  \includegraphics[width=1\linewidth]{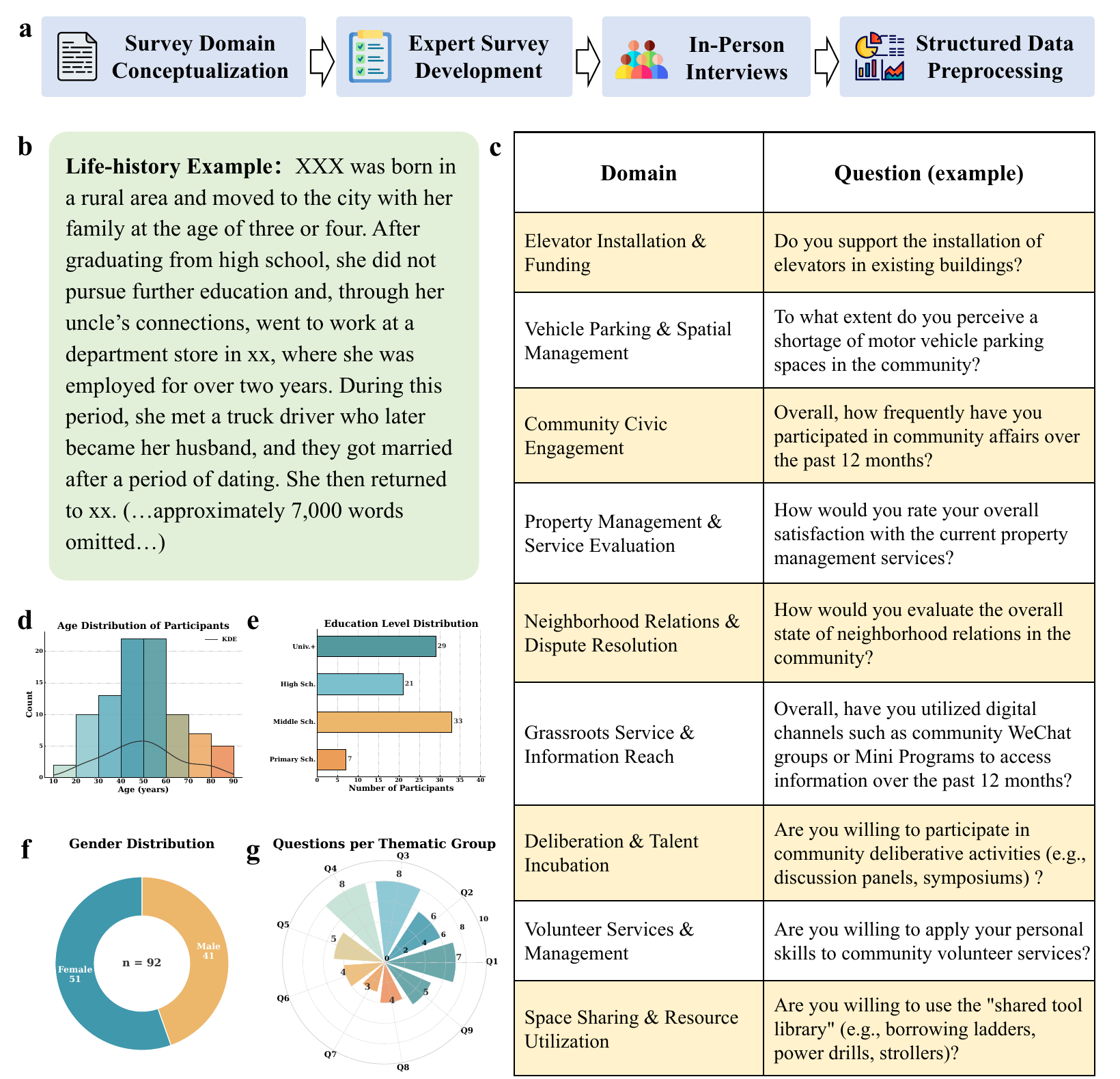}
  \caption{\textbf{A life-history-grounded benchmark for individual-level resident simulation.}
  \textbf{a}, Four-stage data-construction pipeline: survey-domain conceptualization with the local residential committee, expert survey development, in-person semi-structured interviews, and structured data preprocessing.
  \textbf{b}, Excerpt from a representative life-history transcript, illustrating the depth and first-person texture of the narrative material from which each resident's profile is distilled (approximately 7{,}000 words omitted for brevity).
  \textbf{c}, The nine governance domains covered by the 50-question policy-attitude instrument, each illustrated with a representative example question.
  \textbf{d}--\textbf{f}, Demographic composition of the cohort ($n=92$): age (\textbf{d}), education (\textbf{e}) and gender (\textbf{f}).
  \textbf{g}, Number of items per governance domain (Q1--Q9), summing to the 50-question instrument.}
  \label{fig:dataset}
\end{figure*}

\section*{Results}
\subsection*{A life-history-grounded benchmark for individual-level resident simulation}

As motivated in the introduction, community governance requires modelling what specific residents think rather than matching aggregate population distributions. To enable rigorous evaluation of this individual-level resident simulation task, we constructed a benchmark that pairs extensive life-history narratives with structured governance responses from real residents~\cite{park2023generative,argyle2023one,wang2025user,wang2025sociobench,santurkar2023whose}.
The dataset was built in four stages (Fig.~\ref{fig:dataset}a): (i) we worked with the local residential committee to identify the governance issues most salient to residents; (ii) we turned these issues into a structured question instrument; (iii) trained social-science fieldworkers conducted a two-hour interview with each resident, covering both the structured questions and an open-ended life history; and (iv) we transcribed, verified, de-identified, and cleaned the audio recordings, then organized them into life-history profiles paired with each resident's policy-attitude responses. 
Full protocol details, consent forms, and ethics approval are provided in SI~A.

Through this pipeline we recruited 92 long-term residents under stratified sampling targeting balanced coverage across age, gender, and education. Each participant completed a single semi-structured interview of approximately two hours, yielding a rich first-person narrative of that resident's life trajectory and personal values; the full transcribed corpus comprises approximately 1.2 million characters across the 92 residents (mean $\approx$ 13{,}000 characters per resident), and an example is shown in Fig.~\ref{fig:dataset}b. Compared with prior datasets, which are typically \textit{many-respondent, thin-profile} (a lot of respondent, but each with only a thin profile of tens to a few hundred tokens of demographic or persona descriptors~\cite{argyle2023one,santurkar2023whose,hwang2023aligning}), our focus is community governance, where the goal is to model each specific resident and measure per-individual simulation fidelity rather than population-level alignment. Our dataset is correspondingly \textit{few-resident, rich-profile}, deliberately inverting these axes. Faithful individual-level simulation requires a rich per-resident profile, which no amount of additional respondents can substitute for. Our dataset spans the full adult life course (Fig.~\ref{fig:dataset}d), from young adults in their twenties to residents in their eighties, with the largest share in the 40--59 age bracket. The sample includes 51 women and 41 men (Fig.~\ref{fig:dataset}f), and covers the full educational spectrum among residents who disclosed their education: primary school ($n=7$), middle school ($n=33$), high school ($n=21$), and university or above ($n=29$); two residents declined to report their education level (Fig.~\ref{fig:dataset}e). The dataset has two components by design: (i) a \emph{life-history profile} that captures who each resident is, and (ii) a \emph{policy-attitude record} that captures what each resident thinks. The life-history profile is organized into four parts: (P1) basic personal information and growth, (P2) education, work and migration history, (P3) marriage, family and care, and (P4) community interaction and personal values. This four-part division reflects the main categories that sociological life-course research has identified as shaping individual attitudes: personal background (corresponding to P1), accumulated trajectories and role transitions (corresponding to P2), social and familial relationships (corresponding to P3), and subjective interpretation of one's own life (corresponding to P4)~\cite{elder1998life,mayer2009new}. The policy-attitude record is the resident's response to a structured 50-question instrument spanning nine governance domains co-developed with the residential committee to reflect the issues most frequently raised by residents in routine governance interactions (Fig.~\ref{fig:dataset}c): elevator installation \& funding (Q1, 7 items), vehicle parking \& spatial management (Q2, 6), community civic engagement (Q3, 8), property management \& service evaluation (Q4, 8), neighborhood relations \& dispute resolution (Q5, 5), community-level service \& information reach (Q6, 4), deliberation \& talent incubation (Q7, 3), volunteer services \& management (Q8, 4), and space sharing \& resource utilization (Q9, 5) (Fig.~\ref{fig:dataset}g; see SI~A.2.2 for the full 50-item instrument). Items probe both factual circumstances (e.g., whether the building has an elevator) and normative attitudes (e.g., perceived fairness of cost-sharing for elevator installation).

To our knowledge, this dataset constitutes the most extensive corpus to date of paired life-history narratives and structured policy-attitude responses in a community-governance setting, offering a foundation for a range of social-science and computational studies---from analysing how personal life trajectories shape governance attitudes, to studying demographic disparities in policy receptiveness, to serving as a testbed for evaluating LLM alignment with authentic human values across diverse social contexts.

\begin{figure*}[!tb]
  \centering
  \includegraphics[width=0.93\linewidth]{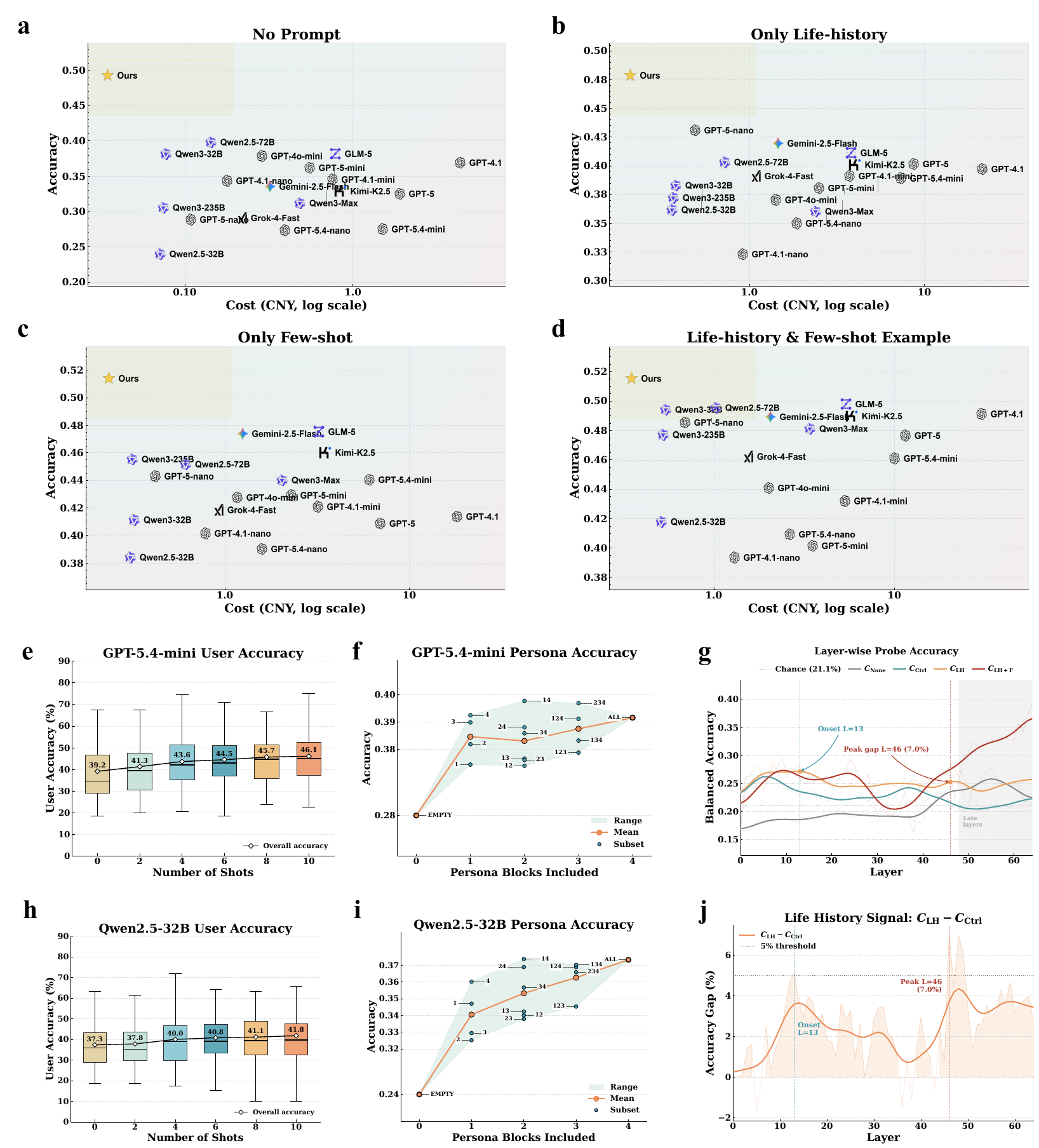}
  \caption{\textbf{Benchmarking 18 mainstream LLMs for individual-resident simulation.}
  \textbf{a}--\textbf{d}, Accuracy vs.\ cost under four prompting strategies (No Prompt, Only Life-history, Only Few-shot, Life-history \& Few-shot); green shading = desirable (low cost, high accuracy); gold band = ``Ours zone'' where \textsc{curriculum-LoRA} (gold star) sits.
  \textbf{e}, \textbf{h}, Per-resident accuracy by number of in-context examples (0--10) for GPT-5.4-mini (\textbf{e}) and Qwen2.5-32B (\textbf{h}); diamonds = overall accuracy.
  \textbf{f}, \textbf{i}, Life-history-component ablation for GPT-5.4-mini (\textbf{f}) and Qwen2.5-32B (\textbf{i}) across all 16 subsets of P1--P4; orange line = mean, band = min--max.
  \textbf{g}, \textbf{j}, Layer-wise probing analysis on Qwen2.5-32B-Instruct: \textbf{g}, probe accuracy under four input conditions (None, Ctrl, LH, LH+F), with shaded band marking the attitudinal inference window; \textbf{j}, life-history signal (C$_{\text{LH}}$ $-$ C$_{\text{Ctrl}}$ gap) across layers, with onset layer and peak gap annotated.}
  \label{fig:bench}
\end{figure*}

\subsection*{Benchmarking mainstream LLMs reveals an unfavourable accuracy--cost frontier}

To construct a high-quality dataset for evaluating heterogeneous LLMs, we selected 60 long-term residents from the participants whose interview responses are information-rich, complete, and unambiguous across the 50-item policy-attitude instrument. Because some residents declined to answer particular items or gave responses that could not be scored unambiguously, invalid cells were excluded, yielding 2{,}291 valid (resident, question, answer) triples. For each resident we then partitioned the valid items into held-out evaluation questions and reference questions; the gold answers to the reference questions are exposed to the model as in-context conditioning, while the evaluation questions remain unseen and are scored by exact match.
We evaluated 18 mainstream LLMs spanning open-source families (Qwen2.5-32B/72B, Qwen3-32B/235B/Max, GLM-5, Kimi-K2.5) and proprietary frontier models (GPT-4.1/4.1-mini/4.1-nano/4o-mini, GPT-5/5-mini/5-nano, GPT-5.4-mini/5.4-nano, Gemini-2.5-Flash, Grok-4-Fast)---each combined with four prompting strategies: \emph{No Prompt}, \emph{Only Life-history}, \emph{Only Few-shot}, and \emph{Life-history \& Few-shot Example}. Within each strategy we further swept the number of in-context few-shot examples ($\{0,2,4,6,8,10\}$). Each resident's life-history was organized into four blocks: basic personal information and growth (P1), education, work and migration history (P2), marriage, family and care (P3), and community interaction and personal values (P4). We exhaustively evaluated all $2^4{=}16$ subsets of these blocks; complete accuracy by model, strategy, and shot count is reported in Fig.~S1.
We present the findings below along five axes: the value of life-history conditioning (Fig.~\ref{fig:bench}a--d, f, i), the internal mechanism by which life-history takes effect (Fig.~\ref{fig:bench}g, j), the diminishing returns of in-context few-shot examples (Fig.~\ref{fig:bench}e, h), the limited role of model capability, and the accuracy--cost trade-off that emerges across configurations (Fig.~\ref{fig:bench}a--d).

\textit{Life-history substantially raises individual-level fidelity.} Without few-shot examples, conditioning on life-history raises accuracy by an average of 5.6\,pp across the 18 LLMs (Fig.~\ref{fig:bench}a vs.\ b), with the largest gains on GPT-5-nano ($29.2\% \to 43.1\%$, $+13.9$\,pp) and Qwen2.5-32B ($24.0\% \to 37.4\%$, $+13.4$\,pp). With few-shot examples, adding life-history still yields an average gain of 3.1\,pp (Fig.~\ref{fig:bench}c vs.\ d), with the largest gains on Qwen3-32B ($41.1\% \to 49.4\%$, $+8.3$\,pp) and GPT-4.1 ($41.4\% \to 49.1\%$, $+7.7$\,pp). Life-history therefore helps with or without few-shot examples. To better understand the role of different life-history components, we examined block-level ablation: Fig.~\ref{fig:bench}f, i shows two representative models, and Fig.~S2 in SI~G reports the full breakdown across all 18 LLMs. Three patterns emerge. First, including all four blocks is consistently a safe choice: most models attain their highest accuracy at the full configuration, and we never observe a subset that strictly dominates it. Second, the marginal benefit of additional blocks varies by model: some saturate after a single block (e.g., GPT-5.4-mini) while others gain monotonically as more blocks are added (e.g., Qwen2.5-32B, GPT-5-nano). Third, at any fixed block count, \emph{which} blocks are included matters: two configurations with the same number of blocks can differ by more than two configurations with different block counts, and a handful of single-block configurations even underperform the no-prompt baseline.

\textit{Layer-wise probing reveals where life-history is encoded inside the model.} The benchmarking results above demonstrate \emph{that} life-history conditioning improves fidelity, but not \emph{how} the model processes biographical information internally. To illuminate the internal mechanism, we conducted a layer-wise probing analysis on Qwen2.5-struct, an open-weight model whose intermediate representations are accessible. We extracted hidden states from each of the model's 64 transformer layers and trained a linear classifier (logistic regression) to predict each resident's policy-attitude answer from the layer's representation alone, under four input conditions: no context (C$_{\text{None}}$), a length-matched control prompt with shuffled text from other residents (C$_{\text{Ctrl}}$), full life-history (C$_{\text{LH}}$), and life-history plus few-shot reference traces (C$_{\text{LH+F}}$). Three findings emerge (Fig.~\ref{fig:bench}g). {First, all conditions start near chance (21.1 \%) at early layers (layers 0--10), confirming that superficial token statistics carry little attitudinal signal.} Second, the C$_{\text{LH}}$ and C$_{\text{LH+F}}$ conditions separate from C$_{\text{None}}$ and C$_{\text{Ctrl}}$ starting around layer 13 (onset layer), with the gap peaking at layer 46 (7.0\,pp)---an \emph{attitudinal inference window} in which the model converts biographical narrative into preference-relevant features; by contrast, C$_{\text{Ctrl}}$, which matches the prompt length but lacks meaningful personal content, shows almost no gain over C$_{\text{None}}$ across the entire depth. Third, the life-history signal (C$_{\text{LH}}$ $-$ C$_{\text{Ctrl}}$; Fig.~\ref{fig:bench}j) first exceeds the 5\% threshold at layer 13 and peaks at layer 46, indicating that the model's middle-to-late layers are where biographical narrative is actively compiled into attitudinal representations. These results provide a mechanistic account of why life-history helps: it supplies the model with narrative material that is progressively distilled into attitudinal representations, a transformation that length-matched control text cannot trigger.

\textit{In-context few-shot examples bring diminishing-return gains.} Adding reference question--answer pairs from the same resident steadily improves accuracy (Fig.~\ref{fig:bench}e, h shows two representative models; the full shot sweep across all 18 LLMs is in Fig.~S1 of SI~G). The $0\to 2$-shot jump is the largest, and each subsequent step adds a smaller increment; the $8\to 10$-shot gain averages roughly 1\,pp across the 18 LLMs. The marginal benefit of additional shots is therefore consistently smaller than that of life-history conditioning.

\textit{Bigger models and reasoning-focused models do not simulate residents better.} First, larger size does not yield higher accuracy: mid-scale open-source models (GLM-5 49.7\%, Qwen2.5-72B 49.5\%, Qwen3-32B 49.4\%) match or beat much larger proprietary systems (GPT-5 47.7\%, GPT-4.1 49.1\%) under Life-history \& Few-shot Example. Second, reasoning-focused models (GPT-5.4-mini, GPT-5.4-nano) rank near the bottom under No Prompt (27.5\% each), well below smaller general-purpose models such as GPT-4o-mini (37.2\%). Resident simulation depends on understanding personal values and attitudes, not on logical reasoning. The bottleneck is therefore not model size or reasoning power, but the lack of individual-level information---which is what life-history provides.

\textit{The best prompting configurations come at much higher cost.} Within each model family, switching from No Prompt to Life-history \& Few-shot Example raises per-call cost by 3--4$\times$ while improving accuracy by at most 10--15\,pp. Across model tiers the gap is even larger: frontier proprietary models cost an order of magnitude more than mid-scale open-source models that come within a few percentage points of them. GPT-4.1 costs CNY 38.12 and GPT-5 costs CNY 13.84 for all evaluation calls, while GLM-5 costs only CNY 6.54 and Qwen2.5-72B only CNY 1.50. The upper-left ``low-cost, high-accuracy'' zone is therefore empty under pure prompting (Fig.~\ref{fig:bench}a--d; per-model API pricing in SI~F).
This cost gap matters for real-world use. One round of policy evaluation easily needs tens of thousands of simulated responses (e.g., a 50-question survey for around 100 residents across several policy options), and iterative refinement multiplies this demand many times over. With frontier prompting, a single round can cost thousands of CNY, far beyond the budget of most local organisations. What matters in practice is therefore not raw accuracy but \emph{accuracy per CNY}: matching the prompting ceiling at a tenth of the cost is what turns an academic demonstration into a deployable tool. We address this accuracy--cost gap next.


\begin{figure*}[!tb]
  \centering
  \includegraphics[width=0.95\linewidth]{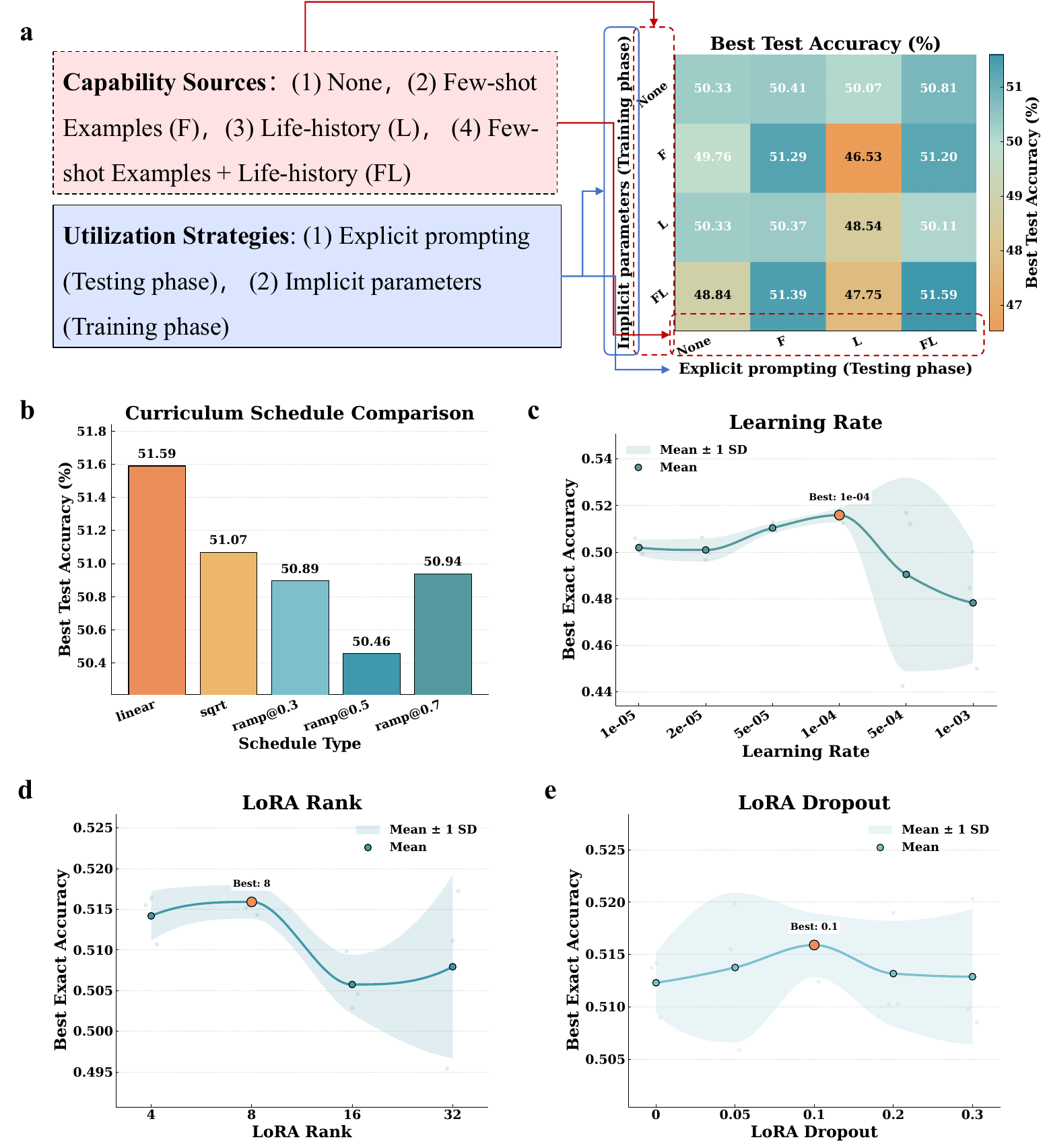}
  \caption{\textbf{\textsc{curriculum-LoRA} consistently improves resident-simulation fidelity.}
  \textbf{a}, Implicit (training phase, y-axis) versus explicit (testing phase, x-axis) use of the two capability sources \textsc{curriculum-LoRA} can leverage---few-shot reference question--answer traces (\textbf{F}) and life-history (\textbf{L}). Each cell reports held-out accuracy under one (train, test) combination, where the four labels denote \textbf{None} (no input), \textbf{F} (few-shot only), \textbf{L} (life-history only), and \textbf{FL} (both).
  \textbf{b}, Comparison of five curriculum schedulers (linear, square-root, and ramp at three end ratios $\rho \in \{0.3, 0.5, 0.7\}$) holding $k_{\min}=2$ and $k_{\max}=10$ fixed; the linear schedule reaches the best held-out accuracy.
  \textbf{c}--\textbf{e}, Hyperparameter sensitivity analysis for \textsc{curriculum-LoRA}: \textbf{c}, learning rate; \textbf{d}, LoRA rank; \textbf{e}, LoRA dropout. Marked points indicate the best configuration.}
  \label{fig:method}
\end{figure*}

\subsection*{\textsc{curriculum-LoRA} shifts the accuracy--cost frontier}

To improve the accuracy--cost trade-off of LLM-based resident simulation, we propose \textsc{curriculum-LoRA}, a parameter-efficient personalization framework that fine-tunes a 7B-parameter base model (\texttt{Qwen2.5-7B-Instruct}) with two lightweight additions to standard LoRA-based supervised fine-tuning. \emph{First}, at every training step we randomly sample a subset of \emph{reference questions}---other governance questions for which the same resident's answer is known---and prepend them, together with the resident's life-history profile, to the target question, so that the model can exploit cross-question consistency in each resident's preferences. \emph{Second}, we apply a curriculum that gradually grows the number of reference questions from $k_{\min}=1$ to $k_{\max}=9$ over the course of training, so that the model first masters the basic life-history-to-answer mapping and only later learns to exploit the richer cross-question context. A complete description, including the loss, scheduling functions and hyperparameter search, is given in Methods (hyperparameter analysis in Fig.~\ref{fig:method}c--e; algorithm pseudocode in SI~D).

\textit{Pareto-dominating the prompting frontier on accuracy and cost.} On accuracy alone, \textsc{curriculum-LoRA} attains 51.6\% exact-match accuracy on the held-out test set (3-seed mean), a 1.9-pp absolute (3.7\% relative) margin over the strongest prompting baseline (GLM-5, 49.7\%); we treat this as a real but modest gain rather than a categorical accuracy advantage. The decisive contribution lies on the cost axis: \textsc{curriculum-LoRA} delivers this fidelity at CNY 0.41 for all evaluation calls---approximately $16\times$ cheaper than GLM-5 (CNY 6.54), $19\times$ cheaper than Kimi-K2.5 (CNY 7.66), $34\times$ cheaper than GPT-5 (CNY 13.84), and $93\times$ cheaper than GPT-4.1 (CNY 38.12) under the same Life-history \& Few-shot Example protocol. On the joint accuracy--cost plane (Fig.~\ref{fig:bench}a--d), \textsc{curriculum-LoRA} (gold star) sits alone in the upper-left ``Ours zone'': no prompting configuration we tested is simultaneously more accurate \emph{and} cheaper. For the deployment scenarios that motivate this work, a fidelity-equivalent solution at order-of-magnitude lower cost is the difference between an academic demonstration and a tool that local organisations can actually run.

\textit{Life-history and few-shot reference traces act synergistically, reinforcing each other at both training and test.}The $4 \times 4$ matrix in Fig.~\ref{fig:method}a maps test accuracy across all 16 combinations of training-time and testing-time inputs from $\{$None, F, L, FL$\}$, where F denotes few-shot reference question--answer traces, L denotes life-history, FL denotes both, and None denotes neither.
\emph{(i) Mismatched train/test inputs can backfire outright.} The matrix's lowest cell is (train:\,F, test:\,L) = 46.53\%, where the model is fine-tuned to expect reference traces at test but is then given life-history alone; fine-tuning is actively harmful here, dropping accuracy by 3.5 pp below the base model on the same test condition, (train:\,None, test:\,L) = 50.07\%. Whatever signal the model is trained to lean on must also be present at test, otherwise fine-tuning hurts rather than helps.
\emph{(ii) The cell (train:\,FL, test:\,FL) = 51.59\% strictly dominates every other cell of the matrix.} All 15 alternative configurations fall below the joint peak, with accuracy ranging from 51.39\% at (train:\,FL, test:\,F), the closest neighbour at 0.20 pp below, down to 46.53\% at (train:\,F, test:\,L), the worst at 5.06 pp below. The peak is reached only when both signals are co-present at both stages: single-signal training falls short even when the test prompt carries the full FL pair (e.g., (train:\,F, test:\,FL) = 51.20\%, (train:\,L, test:\,FL) = 50.11\%), and single-signal testing falls short even when training included both. The two signals are therefore complementary rather than substitutable, and they reinforce each other at training and at test rather than acting only at one.

\textit{Curriculum schedule and hyperparameter sensitivity.} We compared three families of schedules for the curriculum function $g(t/T)$: linear, square-root, and a piecewise ``ramp'' that reaches the maximum at a fraction $\rho$ of training (Fig.~\ref{fig:method}b). The linear schedule attains the highest held-out accuracy (51.59\%), beating square-root (51.07\%) by 0.52 pp and every ramp variant (50.46--50.94\%) by 0.65--1.13 pp. Among ramp variants, a later ramp end ($\rho=0.7$, 50.94\%) is marginally better than earlier ones ($\rho=0.3$ at 50.89\%; $\rho=0.5$ at 50.46\%), but all ramp variants fall below the smoother schedules. Gradually broadening the context across the entire training trajectory is therefore preferred to abruptly jumping to the maximum context size early in training. \textsc{curriculum-LoRA}'s accuracy is most sensitive to learning rate, modestly sensitive to LoRA rank, and barely sensitive to dropout (Fig.~\ref{fig:method}c--e). The optimal learning rate is $10^{-4}$; accuracy is stable for smaller learning rates ($\geq$50\% across $10^{-5}$ to $10^{-4}$) but drops sharply to 44--47\% at $5 \times 10^{-4}$ and above. The best LoRA rank is 8; rank 4 is competitive (51.2\%) while ranks 16 and 32 underperform by 2--3\,pp. Dropout barely matters: across values from 0 to 0.3, accuracy varies by less than 0.2\,pp.

\textit{Generalization to unseen residents and unseen domains.} A deployable simulator must transfer to residents and governance topics not seen during training, since a city has far more residents and policy issues than we can interview. We test this in two experiments. Each varies the size of the training pool and evaluates accuracy on the held-out remainder; the first varies residents, the second varies domains.

\begin{figure}[t]
  \centering
  \includegraphics[width=0.96\linewidth]{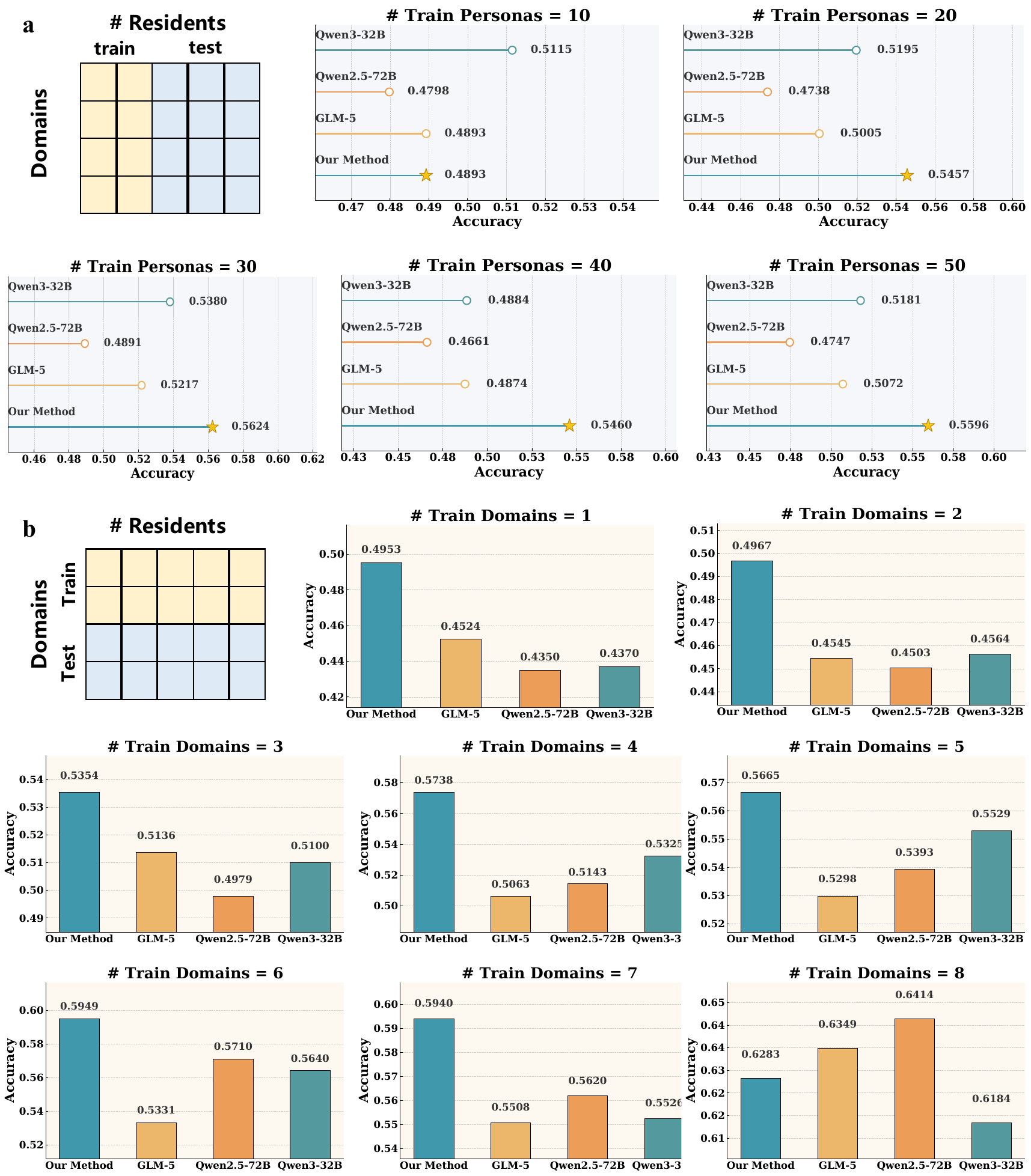}
  \caption{\textbf{Generalization to unseen residents (\textbf{a}) and unseen governance domains (\textbf{b}).} \textsc{curriculum-LoRA} (gold star) is compared against three Life-history~\&~Few-shot Example baselines (Qwen3-32B, Qwen2.5-72B, GLM-5) across training sizes $\{10,20,30,40,50\}$ residents (\textbf{a}) and $\{1,\dots,8\}$ domains (\textbf{b}). Top-left schematics indicate the train/test split.}
  \label{fig:generalization}
\end{figure}

\textbf{Across residents.} We trained \textsc{curriculum-LoRA} on subsets of $\{10, 20, 30, 40, 50\}$ residents and evaluated on the held-out remainder (Fig.~\ref{fig:generalization}a). On average across the five training set sizes, \textsc{curriculum-LoRA} is about 2.6 pp higher than the best baseline. The gap is smallest at train=10 ($-$2.2 pp; \textsc{curriculum-LoRA} matches GLM-5 at 48.9\% and is below Qwen3-32B at 51.2\%) and largest at train=40 ($+$5.8 pp). From train=20 onwards it overtakes every baseline (54.6\% vs 50.0\% for GLM-5 and 52.0\% for Qwen3-32B at train=20), peaks at 30 residents (56.2\%) and plateaus above 54\% out to 50.
Looking at how accuracy changes as the training set grows from 10 to 50 residents makes the difference clearer: \textsc{curriculum-LoRA} gains 7.0 pp, while every prompting baseline barely moves (GLM-5 $+$1.8 pp, Qwen3-32B $+$0.7 pp, Qwen2.5-72B $-$0.5 pp). The contrast exposes a structural advantage of fine-tuning over prompting: each test instance is scored from the target resident's own life-history and reference answers alone, so prompting baselines have no mechanism to consume the other residents in the training pool---changing the train-pool size only redraws the held-out test set, and the small fluctuations seen across pools are test-set noise rather than learning. The LoRA adapter, by contrast, internalises every training resident into its parameters during fine-tuning, so each in-depth interview adds directly to model knowledge regardless of base-model capacity. The complementary insight lies on the data side---an early plateau at 30 residents (one-third of the cohort) implies the life-history-to-attitude mapping is largely \emph{shared} at the population level, so a moderately-sized cohort already covers the latent dispositional structure. In practice, this is the operative scaling regime: collecting hundreds of in-depth interviews per locality is unnecessary, while collecting only a handful is insufficient (see Fig.~S3 for all 18 baselines).

\textbf{Across domains.} We then drew \emph{random} subsets of $\{1, 2, \dots, 8\}$ of the nine governance domains for training and evaluated on the held-out remainder (Fig.~\ref{fig:generalization}b); the random-draw protocol prevents any single ``easy'' or ``hard'' domain from dominating the result at small subset sizes. Averaged across the eight train-domain counts, \textsc{curriculum-LoRA} sits $\approx$2.5 pp above the strongest baseline at each count, with per-count gains as high as 4.3 pp; it outperforms every baseline (GLM-5, Qwen2.5-72B, Qwen3-32B) at counts 1 to 7, and only at 8 training domains does Qwen2.5-72B narrowly edge out \textsc{curriculum-LoRA} (64.1\% vs 62.8\%). The few-domain regime is the more deployment-relevant one: in practice, communities are far more likely to have collected interview answers on one or two governance topics than on all nine, so generalising from a small set of training topics to many unseen ones is exactly the regime that determines whether the simulator is usable in practice. For example, trained on just one of the nine governance domains, \textsc{curriculum-LoRA} attains 49.5\% on the eight unseen domains, 4.3 pp above the strongest baseline. In short, the adapter learns a generalisable mapping from life-history plus a few prior answers to new attitudes, not per-domain patterns; the curriculum then keeps this skill robust to whatever reference-set size a deployment provides (see Fig.~S4 for all 18 baselines).

\subsection*{A closed-loop system for policy simulation and optimization}

\begin{figure}[t]
  \centering
  \includegraphics[width=0.98\linewidth]{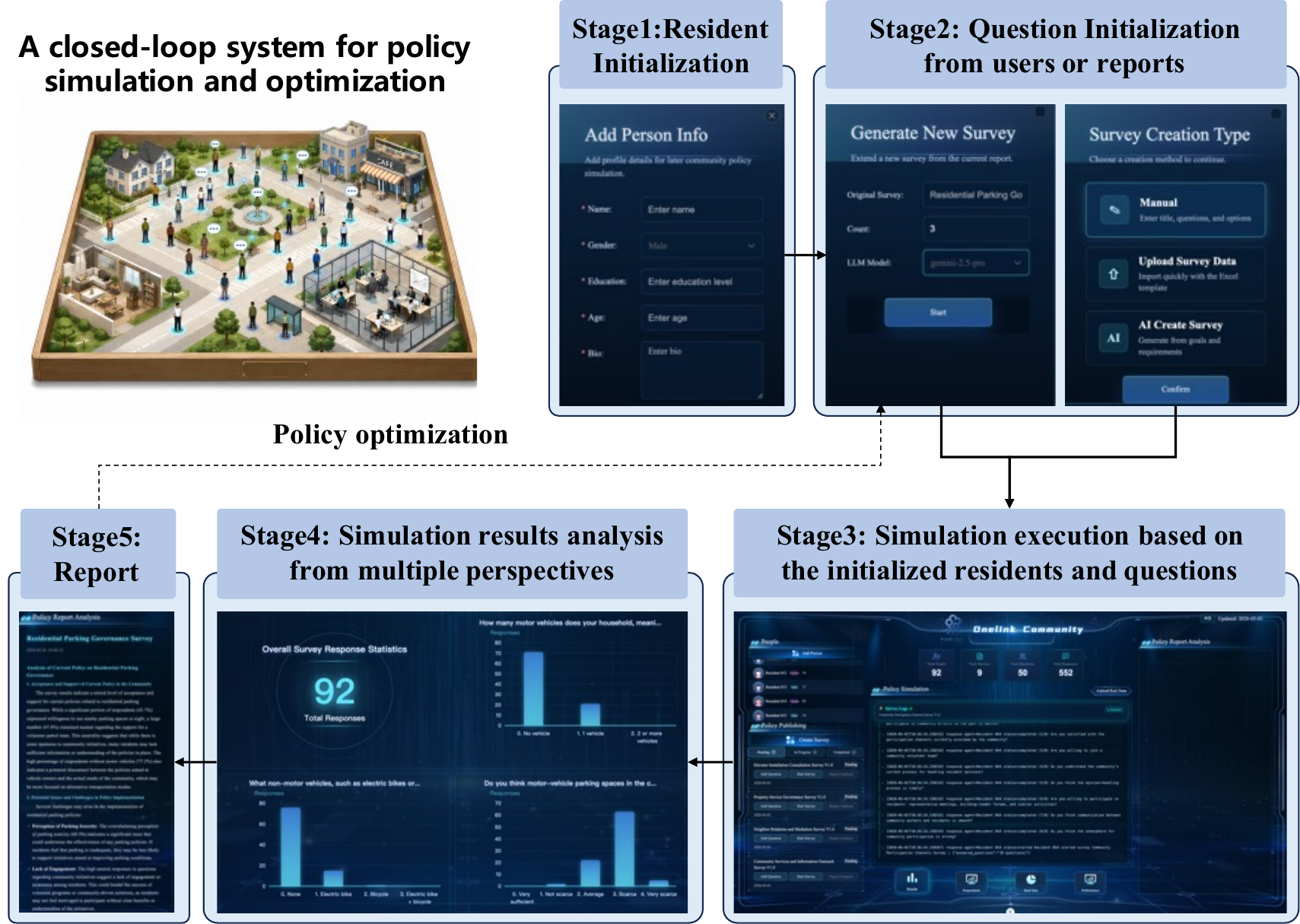}
  \caption{\textbf{A closed-loop, end-to-end platform for policy simulation and optimization.} The deployed \textsc{Onelink-Community} system (\url{http://39.107.76.51:9882/}) operationalizes the Silicon-Society pipeline as five interconnected stages: (i) \emph{Resident initialization}, where life-history-grounded profiles are imported or authored; (ii) \emph{Question initialization from users or reports}, supporting three interchangeable authoring modalities---\textbf{Manual}, structured \textbf{Template} upload, and \textbf{AI-generated} drafting from a stated research goal; (iii) \emph{Simulation execution}, in which curriculum-LoRA resident agents respond at scale and progress is streamed in real time through the community dashboard; (iv) \emph{Multi-perspective results analysis}, presenting aggregate response statistics together with per-question and per-cohort breakdowns; and (v) an automatically synthesized \emph{Simulation report} that diagnoses acceptance, frictions, and equity concerns of the proposed policy. The dashed \emph{Policy optimization} arrow closes the loop: findings re-enter as revised policies for iterative pre-deployment refinement.}
  \label{fig:system}
\end{figure}

To turn the calibrated model into a usable tool for community governance, we built and publicly deployed \textsc{Onelink-Community}, an end-to-end policy-attitude simulation system (Fig.~\ref{fig:system}; live at \url{http://39.107.76.51:9882/}). The system covers the full governance-evaluation workflow in a single closed loop. It is model-agnostic: any LLM can serve as the inference backbone, from general-purpose frontier models to our \textsc{curriculum-LoRA}-calibrated model. Practitioners therefore choose the fidelity--cost operating point that fits their use case. Every stage of our methodology is exposed as an interactive module, from life-history profiles to \textsc{curriculum-LoRA} inference to multi-perspective analytics, so domain practitioners can operate the system end-to-end without machine-learning expertise.

\textit{Operational flow.} The system is structured as five sequential stages threaded by a single feedback loop, so an operator can move from a blank policy idea to a comparable, evidence-based revision without leaving the platform. \textbf{Stage~1 -- Resident initialization.} Operators populate the virtual community by importing existing resident profiles or by entering new ones through the \textit{Add Person Info} form (name, gender, education, age, free-form biography); each profile becomes a persistent resident reused across all subsequent simulations. \textbf{Stage~2 -- Question initialization.} Policy probes are authored in three interchangeable modalities sharing a common schema: \textbf{Manual} authoring of survey title, questions, and options; structured \textbf{Template} upload via an Excel workbook for comparability across studies; and \textbf{AI-generated} drafting in which the operator supplies only a research goal, a target sample size, and an optional prompt and a backbone LLM produces a domain-appropriate questionnaire that the operator may further edit. The same interface also accepts a \emph{prior simulation report} as input, which is the entry point of the optimization loop. \textbf{Stage~3 -- Simulation execution.} Once a survey is created, the dashboard dispatches it to all residents and runs each resident agent in turn; every survey carries three lifecycle controls---\textit{Pending}, \textit{In-Progress}, \textit{Completed}---through which the operator starts, monitors, and closes the run, while a live event log streams each respondent's status. \textbf{Stage~4 -- Results analysis.} Completed simulations are summarized in an interactive analytics view that reports the overall response volume and the answer distribution of each question, giving operators an at-a-glance read of where the simulated community concentrates, splits, or hesitates. \textbf{Stage~5 -- Simulation report and policy optimization.} For every completed run, the system synthesizes a structured \textit{Policy Report} that diagnoses acceptance, anticipated frictions, equity concerns, and concrete revision suggestions, with each claim grounded in the underlying response distribution; the dashed feedback arrow in Fig.~\ref{fig:system} closes the loop, since the report itself becomes a first-class input to Stage~2, enabling iterative pre-deployment refinement before any real-world rollout.

Taken together, these five stages form an end-to-end, publicly deployed pipeline that couples profile-grounded LLM residents, scalable simulation orchestration, multi-perspective analytics, and report-driven policy iteration inside a single web application. The completeness of this loop is what allows the empirical results of the preceding sections to be operationalized as a governance workflow rather than an offline benchmark: practitioners can probe, compare, and refine policies on a silicon community grounded in real human profiles, and only then commit to action in the physical one.

\FloatBarrier
\section*{Discussion}

This work establishes an integrated research foundation for LLM-based resident simulation in community governance, spanning a life-history-grounded benchmark, a systematic accuracy--cost map of 18 mainstream LLMs, the \textsc{curriculum-LoRA} personalization framework, and a closed-loop policy-evaluation system. The following implications emerge.
\emph{A favourable accuracy--cost trade-off makes individual-level resident simulation deployable.} Across 72 model--prompt configurations spanning 18 LLMs, accuracy plateaus near 50\% and is largely insensitive to model scale, profile granularity, or in-context shot count beyond eight. \textsc{curriculum-LoRA} matches the strongest prompting baseline's fidelity (51.6\% vs GLM-5's 49.7\%) at roughly an order of magnitude lower per-call cost ($\sim$$16\times$ cheaper than GLM-5, $34\times$ cheaper than GPT-5, $93\times$ cheaper than GPT-4.1). The practical implication is not that any single LLM is ``winning''---the absolute accuracy difference is modest---but that LLM-based human simulation can now be run at the scale that iterative policy refinement actually requires: tens of thousands of synthetic responses per round, repeated across many policy variants and many iterations, on a budget local administrations can afford. The broader takeaway for the field is that progress in this domain is most usefully measured on the joint accuracy--cost frontier rather than on raw accuracy alone.
\emph{Life-history and few-shot reference traces are synergistic, not substitutable, and both matter at training as well as at inference.} The two information sources \textsc{curriculum-LoRA} leverages---life-history and few-shot reference question--answer traces---contribute most when supplied \emph{together} during fine-tuning. Training without any context leaves the model unable to make full use of either signal at test; training only on life-history barely outperforms this baseline; and training only on reference traces still misses the global maximum by 0.30\,pp. The peak (51.59\% at (train:\,FL, test:\,FL)) is reached only when the model has been exposed to both signals during fine-tuning and reference traces are also activated as the test prompt. Reference traces are therefore not merely an inference-time activator: their inclusion at training time is what teaches the model to integrate cross-question consistency with biographical context, and life-history adds a synergistic lift only when paired with this training-time exposure. Our layer-wise probing analysis provides a mechanistic account of this synergy: life-history narrative is progressively distilled into attitudinal representations within a localised middle-to-late-layer window (onset at layer 13, peak gap at layer 46 in Qwen2.5-32B-Instruct), a transformation that length-matched control text cannot trigger, explaining why rich biographical context is indispensable for individual-level simulation. The general design lesson for LLM-based human simulation is to treat individual evidence (biographical) and task-specific signal (reference traces) as complementary inputs that should be jointly absorbed during fine-tuning, rather than as substitutable inputs that only differ in the stage at which they enter the model.
\emph{Faithfully simulating subjective, value-laden preferences remains the open methodological frontier.} Even with rich life-history conditioning and curriculum-based personalization, accuracy plateaus near 50\%, suggesting fundamental limits on how well current LLMs can infer subjective preferences from biographical context. This is precisely the regime that matters most for community-level policy: governance decisions about parking allocation or cost-sharing for elevator installation hinge on residents' subjective preferences, not on their reported circumstances. Closing this gap is, in our view, the single most important methodological frontier for the field, and likely requires modelling advances that go beyond the next-token formulation we use here.
\emph{A sociological explanation of why life-history data is informative about individual attitudes.} Sociological research on the life course holds that individual attitudes are not static dispositions but form gradually through accumulated experiences, role transitions, and social relationships, with each person continually re-interpreting earlier events in light of current circumstances~\cite{caspi1989continuities,elder1998life,mayer2009new}. The same cost-sharing proposal for elevator installation, for instance, can produce opposite answers from two residents of the same age: one agrees, reading the proposal through a lifetime of pooling resources with neighbours; the other declines, reading it through years of paying for everything alone. More fundamentally, the narrative through which a person tells their own life is not merely a window onto their attitudes but part of how those attitudes are constituted; the transcript therefore encodes dense, individual-level attitudinal signal that no demographic descriptor can carry. These embedded signals enable large language models, trained on narrative text, to predict an individual resident's attitudes directly from their life-history corpus.

Several limitations bound the present work and define its near-term agenda.
\emph{From a single empirical case study to multi-locale instantiation.} The contribution of this work is twofold: a methodological one---a benchmark protocol, a prompting-baseline evaluation framework, and the \textsc{curriculum-LoRA} personalization method---and an empirical one---a single, rigorously-constructed instantiation in one urban community. The methodological contribution is intentionally locale-agnostic: the same benchmark design, prompting strategies, and training procedure can be applied wherever comparable life-history interviews can be paired with structured policy-attitude probes. The empirical instantiation, by contrast, is necessarily local; community-governance practices, cultural norms, and resident expectations vary across regions, and we do not claim that the specific accuracy ceilings we report transfer numerically to other settings. We expect, however, the qualitative patterns to generalise: that prompting alone hits a ceiling far below what policy use requires, that this ceiling is approached only at substantial cost premium, and that targeted parameter-efficient personalization shifts the accuracy--cost frontier. Replicating the methodology in additional governance settings is an immediate next step that would test these qualitative claims and broaden the empirical base of the field.
\emph{From static snapshots to living simulations.} Our dataset captures residents at a single point in time, yet attitudes evolve with life events, policy changes, and social shifts. The deeper ambition is to build living simulations that update as communities change---through periodic re-interviews, integration with public data streams, or LLM-assisted conversational agents that can continuously enrich resident profiles. Such a system would move beyond retrospective analysis toward genuine anticipatory governance.
\emph{From simulation to human--AI collaborative governance.} Our current system simulates resident attitudes toward pre-designed policy probes, but the longer-term potential lies in a more interactive paradigm in which simulated residents first surface candidate concerns, real residents then validate and refine these findings through targeted engagement, and the system iteratively learns from this feedback loop. Such a human--AI collaborative model could reshape how governance decisions are made---not by replacing citizen participation, but by making it more focused, efficient, and informed. The simulator acts as a lens that helps decision-makers ask better questions of real communities, rather than answering on their behalf.
\emph{A broader vision.} The framework presented here---empirical profiling, systematic benchmarking, and personalized calibration on a joint accuracy--cost frontier---extends naturally beyond community governance. Any domain where understanding diverse human responses at scale matters, including public health, education, and urban planning, can adopt the same paradigm. More broadly, our results suggest that as LLMs grow more capable of modelling individual humans, the social sciences will play an increasingly central role in guiding how these capabilities are developed, validated, and deployed responsibly---and that the cost-aware framing of model deployment, not raw capability alone, will determine which of these capabilities actually reach the communities they are meant to serve.

\section*{Methods}

\subsection*{Dataset construction}

\textit{Design rationale: depth over breadth.} Community governance turns on what \textit{specific} residents think, not on aggregate population distributions: a single ground-floor resident's objection to cost-sharing can block an elevator installation that the rest of the building has agreed to fund. Faithful simulation in this setting must therefore be individual-level, conditioned on each resident's own life trajectory. The dataset is correspondingly designed \textit{few-resident, rich-profile} rather than \textit{many-respondent, thin-profile}. Prior LLM-based human-simulation benchmarks pair many respondents with brief persona or demographic descriptors of tens to a few hundred tokens each, supporting population-level alignment studies but limiting what can be learned about how rich individual context shapes individual-level prediction. We invert these axes: 92 residents, each with $\approx$13{,}000 characters of structured life-history paired with 50 single-choice policy-attitude items. This per-resident depth is the precondition for the questions this paper asks: whether longer life-histories improve individual-level simulation fidelity, which life-history components (P1--P4) drive that improvement, and whether a small fine-tuned model can achieve the same fidelity at much lower cost. None of these questions can be posed on \textit{many-respondent, thin-profile} data, regardless of $N$.

\textit{Recruitment and ethical oversight.} We partnered with the residential committee of a representative urban community used as our empirical case study to recruit long-term residents through stratified sampling targeting balanced coverage across age decade, gender, education level and ethnicity. All participants provided informed written consent prior to interview; participation was voluntary and modestly compensated. The study protocol was approved by the institutional ethics committee of the lead institution.

\textit{Interview protocol.} Each resident participated in a single semi-structured interview of approximately two hours, conducted in Mandarin Chinese by trained social-science fieldworkers in a private setting at the residential-committee office. The interview guide had two complementary parts. The first elicited an open-ended life-history narrative organized around four prompts: (P1) basic personal information and growth, (P2) education, work and migration history, (P3) marriage, family and care, and (P4) community interaction and personal values. This four-part organization operationalizes the categories of trajectory, transition, relational, and subjective-meaning variables identified by life-course research~\cite{elder1998life,hareven1994aging,mayer2009new}. The second administered a structured 50-question instrument covering nine governance domains co-developed with the residential committee: elevator installation \& funding (Q1, 7 items), vehicle parking \& spatial management (Q2, 6), community civic engagement (Q3, 8), property management \& service evaluation (Q4, 8), neighborhood relations \& dispute resolution (Q5, 5), community-level service \& information reach (Q6, 4), deliberation \& talent incubation (Q7, 3), volunteer services \& management (Q8, 4), and space sharing \& resource utilization (Q9, 5). 

\textit{Transcription and quality control.} All interviews were audio-recorded and transcribed verbatim. Transcripts were reviewed by a second annotator and corrected against the audio. Personally identifying information (names of individuals, building addresses, workplaces) was removed prior to model use. The final corpus contains 92 residents and approximately 1.2 million characters of life-history narrative (two anonymised full-length sample transcripts in SI~B).

\textit{Question--answer pairs.} For each (resident, question) cell we extracted the resident's selected option and an attitudinal label (positive, neutral, negative) where applicable. Cells where the resident declined to answer or the question was inapplicable were marked invalid and excluded, leaving 2{,}291 valid single-choice (resident, question, answer) triples for training and evaluation.

\subsection*{Benchmarking protocol for prompting baselines}

We benchmarked 18 mainstream LLMs spanning open-source and proprietary frontier systems: \textit{Qwen} (Qwen2.5-32B/72B, Qwen3-32B/235B/Max), \textit{GLM} (GLM-5), \textit{Kimi} (Kimi-K2.5), \textit{OpenAI} (GPT-4.1, GPT-4.1-mini, GPT-4.1-nano, GPT-4o-mini, GPT-5, GPT-5-mini, GPT-5-nano, GPT-5.4-mini, GPT-5.4-nano), \textit{Google} (Gemini-2.5-Flash) and \textit{xAI} (Grok-4-Fast). Each model was queried through its official API with temperature set to 0 to ensure deterministic decoding, except Kimi-K2.5, whose API only allowed temperature 0.6.

For every model we evaluated four prompting strategies that isolate the contribution of life-history and prior question--answer context: \emph{Zero-shot} (question only), \emph{Life-history} (with the life-history profile), \emph{Few-shot} (with reference question--answer pairs only), and \emph{Life-history \& Few-shot} (both).
Combining 18 models with 4 strategies yields 72 prompting configurations. Within each strategy we further swept (i) the number of in-context examples in $\{0,2,4,6,8,10\}$ for the three leading models, and (ii) the subset of life-history parts included in the prompt across all 16 combinations of $\{$P1, P2, P3, P4$\}$ for {two} representative models. All accuracies are reported as exact-match accuracy on the 40-question held-out test set, averaged uniformly over residents.

\textit{Cost accounting.} For each API call we recorded prompt and completion token counts and converted the resulting costs to CNY using each provider's published per-token pricing as of the experiment date; all cost figures reported in the main text are normalized to CNY for all evaluation calls. Because \textsc{curriculum-LoRA} runs on local hardware rather than via an API, we convert its compute consumption into a comparable monetary cost
\begin{equation}
C = \frac{T_{\min}}{60} \times P_{\text{hour}},
\end{equation}
where $T_{\min}$ is the cumulative wall-clock inference time over the full test set (in minutes) and $P_{\text{hour}} = \text{CNY 4.98}$ per GPU-hour is a public-cloud reference rate (AutoDL, \url{https://www.autodl.com/home}); all \textsc{curriculum-LoRA} runs use NVIDIA A800 GPUs. This normalization makes all response costs directly comparable to API-priced baselines (per-model pricing snapshot with average prompt/completion token counts in SI~F; worked cost-calculation example in SI~E).

\subsection*{\textsc{curriculum-LoRA}}

\textit{Task formulation.} We frame resident simulation as persona-conditioned policy-attitude prediction. Let $p_u$ denote the textual life-history profile of resident $u$, and let $q_i = (x_i, \mathcal{O}_i)$ denote the $i$-th policy-attitude question, where $x_i$ is the question text and $\mathcal{O}_i = \{o_{i,0}, \dots, o_{i,C_i-1}\}$ is the set of discrete answer options. For each (resident, question) pair $(u,i)$ the goal is to predict the resident's answer $y_{u,i} \in \{0, \dots, C_i-1\}$. The 50 single-choice questions are partitioned at the question level---uniformly across residents---into a 10-question reference set $\mathcal{Q}^{\mathrm{ref}}$ and a 40-question held-out evaluation set $\mathcal{Q}^{\mathrm{eval}}$. 

\textit{Reference-conditioned prompts.} \textsc{curriculum-LoRA} conditions the model not only on $p_u$ and $q_i$ but also on a subset of reference questions $\mathcal{S}_{u,i} \subseteq \mathcal{Q}^{\mathrm{ref}}$ together with the resident's known answers to those reference questions, exposing cross-question consistency in resident preferences. During training, if the target question is itself in the reference set, we exclude it from the conditioning block by sampling $\mathcal{S}_{u,i} \subseteq \mathcal{Q}^{\mathrm{ref}} \setminus \{q_i\}$. The chat-style prompt
\begin{equation}
m_{u,i} = \mathcal{T}\!\left(p_u, q_i, \{(q_j, y_{u,j})\}_{j \in \mathcal{S}_{u,i}}\right)
\end{equation}
contains a system instruction asking the model to return the option index, followed in order by the life-history profile, the reference question--answer block, and the target question. Sampling is restricted to questions for which resident $u$ has a valid recorded answer. Full chat template and a worked-out instance are given in SI~C.

\textit{Curriculum over context size.} At training step $t$ of $T$ total update steps, we sample $k_t$ reference questions per training example, where
\begin{equation}
k_t = k_{\min} + \left\lfloor (k_{\max} - k_{\min}) \, g\!\left(\frac{t}{T}\right) \right\rfloor.
\end{equation}
We implement three families of scheduling functions:
\begin{equation}
g(r) =
\begin{cases}
r, & \text{linear},\\[2pt]
\sqrt{r}, & \text{sqrt},\\[2pt]
\min(1, r/\rho), & \text{ramp at end-fraction } \rho \in (0,1].
\end{cases}
\end{equation}
We use $k_{\min}=1$ and $k_{\max}=9$ throughout. Online random sampling of reference subsets exposes the model to diverse context combinations within and across epochs (reducing overfitting to any fixed prompt template), while the curriculum gradually broadens the context so that the model first masters the basic life-history-to-answer mapping before exploiting richer cross-question structure. 

\textit{Training objective.} Fine-tuning is performed only on observed triples from the 10-question reference split,
\begin{equation}
\mathcal{D}_{\mathrm{train}}
=
\{(u,i,y_{u,i}) : q_i \in \mathcal{Q}^{\mathrm{ref}},\ y_{u,i}\ \text{is observed}\}.
\end{equation}
For each training target $q_i \in \mathcal{Q}^{\mathrm{ref}}$, the reference subset $\mathcal{S}_{u,i}$ is sampled from $\mathcal{Q}^{\mathrm{ref}} \setminus \{q_i\}$, so the target answer itself is never included in the prompt. The 40-question held-out split $\mathcal{Q}^{\mathrm{eval}}$ is excluded from parameter updates and used only for evaluation. The model is fine-tuned with a causal language-modeling loss masked to assistant response tokens only:
\begin{equation}
\mathcal{L} = -\sum_{(u,i,y_{u,i})\in\mathcal{D}_{\mathrm{train}}} \log p_\theta\!\left(y_{u,i}, \texttt{eos} \mid m_{u,i}\right),
\end{equation}
where $\theta$ are the LoRA adapter parameters added to a frozen base model. LoRA adapters are applied to both attention and feed-forward projection layers; rank/dropout sensitivity is shown in Fig.~\ref{fig:method}d--e. \textit{Inference.} For each evaluation target $q_i \in \mathcal{Q}^{\mathrm{eval}}$, we set $\mathcal{S}_{u,i} = \mathcal{Q}^{\mathrm{ref}}$, using all 10 reference questions as conditioning context. We then predict via constrained classification: a single forward pass scores logits at the final position, and we return
\begin{equation}
\hat{y}_{u,i} = \arg\max_{v \in \{0, \dots, C_i-1\}} p_\theta(v \mid m_{u,i}).
\end{equation}
This avoids unnecessary autoregressive generation and yields deterministic, efficient predictions for discrete-option policy-attitude items.

\subsection*{Training and hyperparameter selection}

We use \texttt{Qwen2.5-7B-Instruct} as the base model and perform a grid search over learning rate $\{1{\times}10^{-5}, 2{\times}10^{-5}, 5{\times}10^{-5}, 1{\times}10^{-4}, 5{\times}10^{-4}, 1{\times}10^{-3}\}$, gradient-accumulation steps $\{4, 8\}$, LoRA rank $\{4, 8, 16, 32\}$, LoRA dropout $\{0, 0.05, 0.1, 0.2, 0.3\}$, and curriculum scheduler $\{\text{linear}, \text{sqrt}, \text{ramp}\}$, totalling 720 configurations $\times$ 3 random seeds $= 2{,}160$ training runs. The best configuration is selected by held-out test accuracy averaged over seeds. Other settings are fixed: 3 epochs, training batch size 1, evaluation batch size 10, max input length 2048, warmup ratio 0.1, AdamW with cosine learning-rate decay and gradient clipping, \texttt{bfloat16} mixed precision, NVIDIA A800 GPUs. Hyperparameter analysis is shown in Fig.~\ref{fig:method}c--e; see SI~D for algorithm pseudocode.

\bibliographystyle{unsrt}
\bibliography{references}






\clearpage
\setcounter{figure}{0}
\renewcommand{\figurename}{Extended Data Figure}
\setcounter{table}{0}
\renewcommand{\tablename}{Extended Data Table}


\end{document}